\documentstyle[12pt]{article}
\def\al{\alpha}
\def\be{\beta}
\def\ga{\gamma}
\def\de{\delta}
\def\ep{\epsilon}

\def\et{\eta}

\def\la{\lambda}

\def\si{\sigma}

\def\ta{\tau}

\def\ph{\phi}

\def\ch{\chi}
\def\ps{\psi}

\def\Ga{\Gamma}
\def\De{\Delta}

\def\La{\Lambda}

\def\Up{\Upsilon}
\def\Ph{\Phi}

\def\cl{{\cal L}}
\def\fr#1#2{{{#1} \over {#2}}}
\def\prt{\partial}

\def\vev#1{\langle {#1}\rangle}
\def\bra#1{\langle{#1}|}
\def\ket#1{|{#1}\rangle}

\def\half{{\textstyle{1\over 2}}}
\def\frac#1#2{{\textstyle{{#1}\over {#2}}}}

\def\lsim{\mathrel{\rlap{\lower4pt\hbox{\hskip1pt$\sim$}}
    \raise1pt\hbox{$<$}}}
\def\gsim{\mathrel{\rlap{\lower4pt\hbox{\hskip1pt$\sim$}}
    \raise1pt\hbox{$>$}}}
\def\sqr#1#2{{\vcenter{\vbox{\hrule height.#2pt
         \hbox{\vrule width.#2pt height#1pt \kern#1pt
         \vrule width.#2pt}
         \hrule height.#2pt}}}}

\def\Re{\hbox{Re}\,}
\def\Im{\hbox{Im}\,}

\def\z{{\bf\hat z}}

\newcommand{\beq}{\begin{equation}}
\newcommand{\eeq}{\end{equation}}
\newcommand{\bea}{\begin{eqnarray}}
\newcommand{\eea}{\end{eqnarray}}
\newcommand{\rf}[1]{(\ref{#1})}

\renewenvironment{thebibliography}[1]
 { \rm
   \begin{list}{\arabic{enumi}.}
    {\usecounter{enumi} \setlength{\parsep}{0pt}
     \setlength{\itemsep}{3pt} \settowidth{\labelwidth}{#1.}
     \sloppy
    }}{\end{list}}

\begin{document}
\titlepage

\begin{flushright}
{IUHET 265\\}
{hep-ph/9501341\\}
{November 1993\\}
{updated April 1994\\}
\end{flushright}
\vglue 1cm

\begin{center}
{{\bf CPT, STRINGS, AND MESON FACTORIES
\\}
\vglue 1.0cm
{V. Alan Kosteleck\'y$^a$ and Robertus Potting$^b$\\}
\bigskip
{\it $^a$Physics Department\\}
\medskip
{\it Indiana University\\}
\medskip
{\it Bloomington, IN 47405, U.S.A.\\}
\vglue 0.3cm
\bigskip
{\it $^b$U.C.E.H.\\}
\medskip
{\it Universidade do Algarve, Campus de Gambelas\\}
\medskip
{\it 8000 Faro, Portugal\\}

\vglue 0.8cm
}
\vglue 0.3cm

\end{center}

{\rightskip=3pc\leftskip=3pc\noindent
Spontaneous breaking of CPT is possible in string theory.
We show that it can arise at a level
within reach of experiments at meson factories currently
being built or designed. For $\phi$, $B$, and $\tau$-charm
factories, we discuss the likely experimental string
signatures and provide estimates of the bounds that
might be attained in these machines.

}

\bigskip\bigskip

\begin{center}
\it
Accepted for publication in Physical Review D
\rm
\end{center}

\vfill
\newpage

\baselineskip=20pt
{\bf\noindent I. INTRODUCTION}
\vglue 0.4cm

To be generally accepted,
any new theory purporting to describe successfully a physical system
must ultimately satisfy two criteria.
First,
it must be viable,
in the sense that it must
provide a physically accurate model for observables
while maintaining internal consistency.
Second,
it must be falsifiable,
i.e.,
it must predict new effects that can be tested experimentally.

String theory provides an ambitious framework within which
to seek a consistent quantum theory of gravity incorporating
all fundamental interactions.
The issue of the viability of string theories has
been the subject of much research and appears promising.
However,
progress on the issue of falsifiability has been slower,
since it entails addressing the difficult problem
of identifying a potentially observable
effect in string theory
that cannot occur in a particle theory.

Although such stringy effects are likely to exist
because strings are qualitatively different
from particles,
the search for an experimentally testable stringy effect
meets several obstacles.
One problem arises from the absence of a satisfactory
and completely realistic string theory.
This can be partially avoided by examining features typical
of string theories and by looking
for ways that stringy effects \it could \rm appear
in present experiments,
without requiring that they \it must \rm occur.
However,
even this weakened notion of falsifiability
faces difficulties because any stringy effects
are likely to be highly suppressed at present energies.
The Planck scale $M_{Pl}$ is the natural string scale,
while present experiments allow access only to energies
of order of the electroweak scale $m_{ew}$.
Stringy effects are therefore likely to be suppressed
by some power of the ratio $m_{ew}/M_{Pl}\simeq 10^{-17}$,
making them difficult to observe.
For most purposes,
at present energies the string appears pointlike
and so presumably is well described in terms of a
four-dimensional renormalizable gauge theory.

One method of minimizing the impact of the suppression
is to seek string properties that
violate an exact symmetry of particle theory,
preferably one that can be measured to a high degree
of precision.
One candidate symmetry of this type
is the discrete symmetry CPT.
Although CPT is known to be a symmetry of
local relativistic point-particle field theory
under a few apparently mild assumptions
\cite{S1,S2,S3,S4,S5,S6,S7},
the extended nature of strings means
that CPT invariance is not apparent
\it a priori \rm in the string context.
In addition,
CPT is indeed accessible to
high-precision tests.
Presently, the best experimental bound on CPT violation
is obtained from observations of the $K \overline K$ system
\cite{C1,C2},
where one figure of merit is
\beq
\fr {m_K- m_{\overline K}} {m_K} \le 5\times 10^{-18}
{}~~.
\label{a}
\eeq
Together,
these features of CPT suggest it could provide
a signature for string theory.

A mechanism by which certain string theories
could spontaneously break CPT invariance is known
\cite{KP1,KP2,KP3}.
A summary of the mechanism is presented in sect.\ II.
The main goal of this paper
is to address the issue of how
stringy CPT breaking might manifest itself
at present energies and how best to attempt
to observe it experimentally.
In sect.\ III,
we consider possible effects of the CPT breaking
in a generic neutral meson-antimeson system,
denoted $P \overline P$.
The possibility of using various meson factories to measure
CPT-violation parameters is discussed in sect.\ IV.
We provide some estimates on the limits that
could be attained in machines currently
being built or considered,
including $\ph$, $B$, and $\ta$-charm factories.
Sect.\ IVA presents features generic to any
$P \overline P$ system,
while sects.\ IVB through IVD discuss
aspects specific to the various mesons.
We summarize in sect.\ V.
Some details of our error analysis are relegated
to an appendix.

Throughout,
we use the symbol $P$ to denote any of
the relevant neutral mesons.
Our notation is based on that of the
classic work on the $K\overline K$ system
by Lee and Wu, ref.\ \cite{leewu}.
We have, however, adopted slightly different phase conventions
for the CP properties of the states $\ket{P}$;
see sect.\ III.

\vglue 0.6cm
{\bf\noindent II. BACKGROUND: CPT AND STRINGS}
\vglue 0.4cm

The possibility that certain string theories could
spontaneously break CPT invariance,
at a level that might be observable in the $K\overline K$
system,
was suggested in ref.\ \cite{KP1}
on the basis of an investigation of CPT
in string field theory.
To make this paper more self-contained,
this section summarizes some of the earlier analysis;
the reader desiring to skip the preliminaries can
proceed directly to section III.
In what follows,
we work at the level of field theory
so that off-shell properties
are correctly incorporated.

The issue of the CPT properties of any given
particle or string theory
can be addressed via several approaches,
at least in principle.
A highly sophisticated approach for particle
field theories is provided by the axiomatic method.
At present, however,
this method is unavailable for strings
because appropriate axioms
have not yet been formulated.
Another general approach is the constructive method,
in which particle field theories are viewed as
built from products of anticommuting irreducible spinors,
with overall CPT properties inferred
from those of the underlying spinors.
This method, too, is not presently practical
for strings.
Both these approaches
remain open areas for future investigation.

An approach that can be applied to individual particle
and string theories exists
\cite{SA}.
The method begins by establishing
suitable C, P, and T transformations
for the free quantum theory
and proceeds by applying these to the interacting
fields by expanding them in a Dyson series,
thereby establishing the CPT properties of the full theory.
Application of this method to string theory
requires some care;
for example,
string interactions modify canonical quantization
so the compatibility of
the initial free-field C, P, and T transformations
must be verified.

In ref.\ \cite{KP1},
this method was applied to
the field theories for the open bosonic string \cite{W}
and the open superstring \cite{PTY,AMZ}.
Some standard assumptions were used
including, for example,
the connection between spin and statistics.
It was shown that the bosonic-string action
is C, P, T, and CPT invariant,
and that previously proposed superstring actions
violate C, P, and T but preserve CPT.
Whereas the superstring P and T violation results from the appearance
of massless chiral fermions in ten dimensions,
the C violation was unexpected.
However,
ref.\ \cite{KP1} presents a modified superstring
action maintaining both C and CPT.
The analyses suggest that,
despite the extended nature of the string
and the corresponding infinite number of particle fields,
string and particle theories are sufficiently alike
that CPT is preserved at the level of the action.

In addition to the CPT properties of the action,
the CPT behavior of the ground state must also
be considered.
Noninvariance of the vacuum then corresponds to
spontaneous CPT violation.
While this is unnatural in conventional particle models,
the situation is less obvious in string theory,
which has an off-shell description
naturally lying in higher dimensions.
If strings are to describe the real world,
the corresponding higher-dimensional Lorentz
invariance \it must \rm be broken,
to at least an approximate four-dimensional one.
Moreover,
the mechanism for breaking Lorentz invariance
should be implicit in the action for the higher-dimensional string.

A natural mechanism of this type does exist in string field theory
\cite{KS1}.
The mechanism also breaks the higher-dimensional
CPT invariance.
It then becomes an open question as to whether
four-dimensional CPT is preserved.
Consider,
for example,
the interaction lagrangian of the field theory for
the open bosonic string.
This contains a term
${\cal{L}}_1=\phi A_\mu A^\mu$,
which controls the coupling between
the tachyon $\phi$ and the massless vector $A_\mu$.
However,
the tachyon vacuum expectation value $\vev \ph$
is driven away from the
origin since the effective potential has a local maximum there.
For the appropriate sign of $\vev \ph$
a similar instability is generated in the
effective potential for $A_\mu$,
and any resulting $\vev {A_\mu} \ne 0$
spontaneously breaks the 26-dimensional Lorentz symmetry.
Since $A_\mu$ changes sign under CPT,
CPT-violating terms appear in the action.
The mechanism is stringy in the sense that particle gauge invariance
excludes terms of the form ${\cal{L}}_1$
in a standard four-dimensional gauge field theory,
whereas they are compatible with string gauge invariance.
Moreover,
the mechanism can involve other Lorentz scalars $S$ and tensors $T$
because string field theories typically have cubic interactions
of the general form $S T \cdot T$.

Since no zeroth-order CPT violation appears
in laboratory experiments,
any observable effects must be suppressed,
perhaps because higher-level fields of Planck-scale mass
are involved.
The ratio $m_l/M_{Pl}\lsim 10^{-17}$
of the low-energy scale $m_l$
to the Planck scale $M_{Pl}$
provides the natural dimensionless quantity governing
the suppression.
Observable effects could arise in a variety of $P \overline P$
systems,
as we discuss below.
For instance,
in the $K \overline K$ system one might expect
\cite{KP1,KP4,KP5}
\beq
\fr {m_K- m_{\overline K}} {m_K} \sim \fr {m_l}{M_{Pl}}
{}~~,
\label{b}
\eeq
which includes a range just below the present bound,
Eq.\ \rf{a}.

We remark here that the ratio $m_l/M_{Pl}$ might be expected
to arise in the context of any gravitational mechanism
that violates CPT.
Even in canonical general relativity
the precepts of the usual CPT theorem may not hold
\cite{hawking,page,wald}
so CPT violation at observable levels might occur
\cite{ellis1,peskin}.
However,
the stringy spontaneous CPT violation we discuss here
arises from a specific mechanism
and has a particular characteristic signature (see below).
A recent discussion of spontaneous CPT violation for strings
in the gravitational context is given in
ref.\ \cite{ellis2}.

\vglue 0.6cm
{\bf\noindent III. STRINGY CPT VIOLATION IN A $P \overline P$ SYSTEM}
\vglue 0.4cm

To make contact with experiment,
we must first connect the potential spontaneous CPT
violation in a string theory to
terms in a four-dimensional low-energy
effective action and then determine the effect
on measurable quantities
\cite{KP4,KP5}.
In the absence of a specific,
satisfactory, and completely realistic string field theory,
it is necessary instead to work within
a generic theoretical framework.
In this spirit,
the present section considers a class of possible terms
and investigates their implications.

We assume that the four-dimensional effective theory
arises from a string theory compactified at the Planck
scale,\footnote{
String theories partially compactified at larger length scales
can produce other string signatures;
see refs.\ \cite{KS2,antoniadis} and references therein.}
leading among other terms to
four-dimensional effective interactions
of the schematic form
\beq
\cl_I \supset \fr {\la}{M_{Pl}^k}~T \cdot \bar\ps \Ga (i\prt )^k \ch
 + {\rm h.~c.}~~.
\label{c}
\eeq
For notational simplicity,
the Lorentz indices in this equation are suppressed.
The field $T$ is a four-dimensional Lorentz tensor,
while
$\ps$ and $\ch$ are four-dimensional fermions,
possibly identical.
The symbol $\Ga$ is used to denote a gamma-matrix structure,
while $(i\prt )^k$ represents a $k$th-order derivative coupling
with the four-dimensional derivative $\prt_\mu$.
By definition,
$k \ge 0$.
Each term in the interaction lagrangian
of the string field theory is a Lorentz scalar,
so $T \cdot \Ga (i\prt )^k$ represents a
spinor matrix with derivative entries.
The coupling $\la$ is dimensionless.
The inverse of the $k$th power of the Planck scale
arises from derivative couplings in the string field
theory and the compactification process,
and ensures the correct overall dimensionality.

Following the mechanism described in sect.\ II,
we suppose that the tensor $T$ acquires
an expectation value $\vev T$.
The interaction terms Eq.\ \rf{c}
then generate terms in the lagrangian quadratic
in the fermion fields,
and hence produce a
tree-level contribution $\De K(p)$ to the
fermion inverse propagator $K(p)$.
Suppressing the Lorentz indices on the
momentum factors,
$\De K(p)$ is given by
\beq
\De K(p) = \fr {\la}{M_{Pl}^k}~\vev T \cdot \Ga p^k
{}~~.
\label{d}
\eeq

To incorporate a measure of the possible suppression,
we write the expectation $\vev T$ as
 \beq
\vev T = t \left( \fr {m_l}{M_{Pl}} \right)^l M_{Pl}
{}~~.
\label{e}
\eeq
Here,
$t$ is a numerical factor incorporating the Lorentz
structure of $T$.
It is tempting theoretically to assume numerical values of
order one for the nonzero components of the product $\la t$,
for all $P$.
However,
this assumption presumably would hold at best only at the
scale at which Eq.\ \rf{c} becomes applicable,
below which running would occur.
Moreover,
it would appear that
any assumption of this type should be treated with
caution in view of the observed differences in the scales
of the Yukawa couplings in the standard model.
These are also trilinear boson-fermion-fermion couplings,
but range from about $10^{-5}$ for the up quark
to about $10^{-1}$ for the bottom quark.

The ratio $(m_l/M_{Pl})^l$
in Eq.\ \rf{e} allows for possible suppression
by a power of the low-energy scale to the Planck scale.
By assumption,
$l\ge 0$.
Since we are interested in the
situation where $T$ has nontrivial Lorentz structure
and incorporates CPT breaking,
it appears on experimental grounds that
a realistic model must have $\vev T \ll m_l$, i.e., $l \ge 2$.
A theoretical argument deducing
that this \it must \rm occur
would require the solution to a hierarchy problem
that evidently lies outside the scope of the present work.
Its resolution
may be related to that of the usual hierarchy problem
since the same scales appear.
Indeed, we are using the experimental fact of
its existence to motivate the
introduction of the small scale $m_l$ in Eq.\ \rf{e}.
In any case,
an understanding of the mechanism responsible for
the appearance of disparate scales in nature
presumably awaits the formulation and study of
a satisfactory, realistic string (field) theory.

For a fermion of mass $m_f$,
the order of magnitude of the
change in the fermion inverse propagator
relative to its size is therefore
\beq
\fr{\De K}{K} \sim \left( \fr {p}{M_{Pl}} \right)^k
\left( \fr {m_l}{M_{Pl}} \right)^{l-1}
\left( \fr {m_l}{m_f} \right)
{}~~.
\label{f}
\eeq
For purposes of comparison with experiment,
we assume the fermions to be observable
so that $m_f \sim m_l$ and $p \ll M_{Pl}$.
Since no zeroth-order CPT violation
has been detected in laboratory experiments,
it follows that $k+l > 1$.
Any associated Lorentz breaking is then small
and has no effects other than the CPT-related
ones discussed below.
With the above assumptions,
the question of the experimental detection
of lowest-order stringy spontaneous CPT violation
reduces to determining
the observable consequences of
terms with $k=0$ and $l=2$.

It can be seen from Eq.\ \rf{e}
that terms of this type
produce effects suppressed by about 17 orders of magnitude,
so their direct experimental detection is difficult.
Instead,
we consider interferometric experiments,
which are the ones of choice for high-precision purposes.
In this paper,
we concentrate on the
interferometric experiments that can be developed
for the $P \overline P$ systems,
where we use $P$ generically to denote one of the neutral mesons.
The fermions $\ps$ and $\ch$ of Eq.\ \rf{c}
can be taken as one or more
component quark fields of the meson $P$.
Since we are interested in the dominant contributions
from the effective interactions in Eq.\ \rf{c},
where needed in what follows
we work in lowest-order perturbation theory.
For example,
this implies the replacement of $\De K(p)$
with its amplitude in the quark wavefunctions.
It also means that terms involving both flavor changes and CPT violation
can be neglected.

To make further progress,
we must consider effects on the $P \overline P$ system
of the changes $\De K$ in the
matrices of the quark inverse propagators.
In principle,
there are many different linearly independent ways
for CPT to be violated by each term of the form \rf{c},
one for each possible CPT-violating expectation of
the tensor $T$.
However,
the $P \overline P$ system acts as an \it energy \rm
(mass) interferometer
(as can be seen from the effective-hamiltonian formalism below),
so the only experimentally detectable effect
involves an energy shift.
For example,
an interaction of the form
$\la_q T_{\la\mu\nu}\bar q\ga^\la\ga^\mu\ga^\nu q$
together with an expectation value
$\vev {T_{000}} = t_{000} (m_l/M_{Pl})^2 M_{Pl}$
provides a direct energy shift
$E \to E + \la_q m_l^2 t_{000}/M_{Pl}$ in the
$q$-quark inverse propagator
$E\ga^0 - \vec p \cdot \vec\ga + \ldots$.
Indirect energy shifts are also possible.
For example,
any term leading to a three-momentum shift
also changes the energy through the dispersion relation.
The combination of such shifts may lead to (partial)
directional dependence of the energy effects.
For simplicity,
we disregard any directional effects in the present work,
and we focus on the experimental detection of the
possible energy shifts.

The energy shifts in the quark propagators
induce corresponding shifts in the mass and decay
matrices for the $P \overline P$ system.
Before proceeding further,
we first introduce the notation and conventions
we use for these matrices,
allowing for CPT violation.
Define the eigenstates
$\ket{P^0}$ and $\ket{\overline {P^0}}$
of the T- and CPT-invariant
part of the full effective four-dimensional action
by
\beq
CP \ket{P^0} = \ket{\overline {P^0}}~~,~~~~
CP \ket{\overline {P^0}} = \ket{P^0}
\quad .
\label{g}
\eeq
In the
$\ket{P^0}$-$\ket{\overline {P^0}}$
state space,
the time evolution of a linear superposition
of $\ket{P^0}$ and $\ket{\overline {P^0}}$
is governed by a two-by-two effective hamiltonian
$\La$,
which can be written in terms of two
hermitian operators $M$ and $\Ga$
called the mass and decay matrices, respectively
\cite{leewu}:
\beq
\La= M - \half i \Ga
\equiv
\pmatrix{\La_{11}&\La_{12}\cr
\La_{21}&\La_{22}\cr}
\quad .
\label{h}
\eeq
The matrix elements can be determined in
a degenerate perturbation series
using the T and CPT violating terms in the full action
as the perturbation.
If CPT invariance holds,
$\La_{11} = \La_{22}$.
In what follows,
it is useful to introduce the parametrization
\beq
i\La=\half \Ga  + i M \equiv
\pmatrix{D+iE_3&iE_1+E_2\cr
iE_1-E_2&D-iE_3\cr}
\quad ,
\label{i}
\eeq
and, for convenience, the quantity
\beq
E^2 = E_1^2+E_2^2+E_3^2
\quad .
\label{k}
\eeq
In the above expressions,
a subscript $P$ distinguishing analogous quantities
in distinct $P\overline P$ systems is understood
but omitted for notational simplicity.

The physical
particles\footnote{
The subscripts are defined
in analogy with the CP properties of the kaon system,
where they refer to the particles with `short' and `long'
lifetimes.
Note, however, that the particle $P_S$ is not
necessarily the shorter-lived state in general.
}
$P_S$ and $P_L$
are represented by the two eigenvectors
$\ket{P_S}$ and $\ket{P_L}$
of the matrix $\La$.
The corresponding eigenvalues $\la_S$ and $\la_L$
are combinations of the masses and lifetimes
of the physical particles,
\beq
\la_S = m_S - \half i \ga_S ~~,~~~~\la_L = m_L - \half i\ga_L
\quad .
\label{n}
\eeq
Note that a subscript $P$ on
each of these quantities is again understood.

In terms of the parametrization of Eq.\ \rf{i},
we find
\bea
i\la_S & = &D+iE~~,~~~~
\ket{P_S}= \fr {(1+\ep_P + \de_P ) \ket{P^0}
+(1-\ep_P -\de_P )\ket{\overline{P^0}}}
{\sqrt{2(1+\vert \ep_P + \de_P \vert^2)}}
{}~~, \nonumber\\
i\la_L & = &D-iE~~,~~~~
\ket{P_L}= \fr {(1+\ep_P - \de_P ) \ket{P^0}
-(1-\ep_P +\de_P )\ket{\overline{P^0}}}
{\sqrt{2(1+\vert \ep_P - \de_P \vert^2)}}
\quad .
\label{j}
\eea
Here, the parameter
\beq
\ep_P = \fr { -iE_2}{E_1 + E}
\quad
\label{l}
\eeq
is a measure of T violation and
\beq
\de_P = \fr {E_3}{E_1 + E}
\quad
\label{m}
\eeq
is a measure of CPT violation.
We have explicitly included the subscript $P$ on the
parameters $\ep_P$ and $\de_P$
to avoid confusion in what follows.
Also, it follows that for each $P$
\bea
E & = & - \half \De m - \frac 1 4  i \De \ga
\nonumber\\
  & = & - \half i a \exp (- i \hat\ph)
\quad ,
\label{o}
\eea
with\footnote{
Note that in general $\De m $ and $\De \ga$ are
not necessarily positive quantities,
since the correspondence between CP eigenstates
and the relative masses and lifetimes of the
physical particles may not be the same as that
in the neutral kaon system from which the notation
is abstracted.
The sign changes that ensue
do not affect the results that follow.
}
\bea
\De m = m_L - m_S
{}~~ & , &~~~~
\De \ga = \ga_S - \ga_L
\nonumber\\
a = (\De m ^2 + \frac 1 4 \De \ga^2 )^\half
{}~~ & , &~~~~
\hat\ph = \tan^{-1} \fr{2 \De m}{\De \ga}
\quad .
\label{oo}
\eea
For the purposes of sect.\ IV,
it is also convenient to define
\bea
m = m_S + m_L
{}~~ & , &~~~~
\ga = \ga_S + \ga_L
\nonumber\\
b = (\De m ^2 + \frac 1 4 \ga^2 )^\half
{}~~ & , &~~~~
\tilde\ph = \tan^{-1} \fr{2 \De m}{\ga}
\quad .
\label{ooo}
\eea

The parameters of interest for CPT studies are the $\de_P$,
introduced in Eq.\ \rf{m} above.
Combining this equation with
Eqs.\ \rf{d} and \rf{e}
and the discussion in the paragraphs following Eq.\ \rf{f}
permits the various $\de_P$ to be expressed in terms of quantities
appearing in the effective low-energy action \rf{c}
and the experimentally accessible parameters
$\De m$, $\De \ga$ and $\hat\ph$.
We next proceed to show this.

Since the CPT-violating term is so small,
it is sufficient to work at lowest order
in perturbation theory in the CPT-violating coupling $\la$
of Eq.\ \rf{c}.
As discussed above,
the only observable effects arise
as energy shifts in the quark inverse propagators.
At this level in perturbation,
terms of the form \rf{c} then contribute
only energy insertions on quark lines
in the $P$ and $\overline P$ mesons.
Moreover,
the CPT violation means that
quark and antiquark insertions of given flavor have
the same magnitude but opposite sign.
For instance,
in the example considered above
with interaction term
$\la_q T_{\la\mu\nu}\bar q\ga^\la\ga^\mu\ga^\nu q$
and expectation value
$\vev {T_{000}} = t_{000} (m_l/M_{Pl})^2 M_{Pl}$,
any $q$-quark line in the meson would receive a perturbative
correction with Feynman weight
$\la_q m_l^2 t_{000}/M_{Pl}$
while a $\bar q$-antiquark line acquires
the same weight but with opposite sign.

The $P$ and $\overline P$ mesons are comprised of valence quarks
lying in a sea of gluons, quarks, antiquarks, and other particles
present in the effective action for the low-energy theory.
Let us first disregard the presence of the sea;
below we show that this is a good approximation
within the level of accuracy needed for our purposes.
Since the $P$ and $\overline P$ are antiparticles,
each energy insertion on a valence-quark/antiquark line in
$P$ (or $\overline P$) has a corresponding contribution
of equal magnitude but opposite sign
in $\overline P$ (or $P$).
The contributions from the valence-quark energy insertions
evidently provide equal-magnitude but opposite-sign
contributions to the two diagonal elements
of the effective hamiltonian $\La$,
given at lowest order in the CPT-violating coupling $\la$
by the expectation of $-\cl_I$
in the $P$-meson wavefunction:
\beq
\De \La_{11} = - \De \La_{22} = - \bra{P^0}\cl_I\ket{P^0}
\quad .
\label{r}
\eeq
Note that contributions to $\De \La_{12}$ and $\De \La_{21}$,
present in principle,
are suppressed
because they involve either
two CPT-violating flavor-changing insertions
or higher-order weak effects along with a first-order CPT-violating
flavor-neutral insertion as above.

In general,
the contribution $h_{q_j}$ to $\De \La_{11}$
from the $j$th valence quark, $j=1,2$,
is given in terms of the corresponding Feynman weight
arising from Eqs.\ \rf{d} and \rf{e}:
\beq
h_{q_j} = r_{q_j} \la_{q_j} t \fr {m_l^2}{M_{Pl}}
\quad ,
\label{ra}
\eeq
where the factors $r_{q_j}$, discussed below,
have been included to compensate for
the use of the valence-quark approximation.
It follows that the shifts in the diagonal elements of the
effective hamiltonian are
\beq
\De \La_{11} = - \De \La_{22} = h_{q_1} - h_{q_2}
\quad .
\label{s}
\eeq
Note that in writing this expression we have implicitly
assumed that for each quark flavor
only one term of the form in $\cl_I$
appears, or at least that one such term dominates.
However,
additional terms can readily be incorporated by summing over the
different $h_{q_j}$ contributions.
Their presence would not affect the order-of-magnitude estimates
and other results presented below.

One possible contribution to the factors $r_{q_j}$
arises because energy insertions could be made on sea quarks.
However,
where the sea quarks occur as same-flavor quark-antiquark pairs,
the two possible first-order CPT-violating contributions cancel
because they have opposite sign.
The only sea-quark contributions therefore arise from
flavor-changing interactions,
which are suppressed.
Whether or not there are CPT-violating terms,
the presence of the quark-gluon sea itself
accounts, for example, for the difference between the
mass of a light meson (kaon) and the sum of the current
masses of its quarks.
For instance,
in the example considered above
with interaction term
$\la_q T_{\la\mu\nu}\bar q\ga^\la\ga^\mu\ga^\nu q$
and
$\vev {T_{000}}\ne 0$,
the expectation
$-\bra{P^0}\cl_I\ket{P^0}$
reduces to the Feynman weight
multiplying the meson-wavefunction expectation
of the $q$-quark number operator.
Disregarding flavor-changing effects,
the latter is just one.
In any event,
although the factors $r_{q_j}$
might be important for a detailed calculation,
they play no significant role for the order-of-magnitude estimates below.
In what follows,
we therefore take these factors to be one for simplicity.

We can now combine the above results.
Setting $E_3 = \De \La_{11}$ and
making for convenience
the approximation $E_1 \approx E$,
which is valid provided T and CPT violation are small
(i.e., $E_2$ and $E_3$ small compared with $E$),
we obtain the expression
\beq
\de_P = i \fr {h_{q_1}-h_{q_2}}
{\sqrt{\De m^2 + \De \ga^2/4}}
e^{i\hat\ph}
\quad .
\label{t}
\eeq
This equation applies to any $P \overline P$ system
for any CPT-violating term of the form $\cl_I$ in Eq.\ \rf{c}.
We remind the reader
that subscripts $P$ are understood on all parameters
on the right-hand side.

Establishing the experimental signature of stringy
CPT violation is at this point reduced to
determining the net contributions
to $\de_P$ from different interaction terms \rf{c}
via Eq.\ \rf{t}.
An important aspect of this process in what follows
is that the energy shifts in the effective hamiltonian
are real.
This is true because the dominant effects from the CPT-violating
terms arise as simple matrix elements of the interaction terms
in Eq.\ \rf{c},
which are real since the fundamental string theory
is hermitian.
Using Eq.\ \rf{t},
it follows that
\beq
\Im \de_P = \pm ~{\rm cot} \hat\ph ~\Re \de_P
\quad ,
\label{tt}
\eeq
a result that we use in sect.\ IV.

A range of plausible values of $|\de_P|$ can be
identified from Eq.\ \rf{t}.
The key issues are the sizes of the light mass
scale $m_l$ and the factors $r_q \la_q t$
appearing in Eq.\ \rf{s}.
The natural value of $m_l$ lies between the mass $m_P$
of the $P$ meson and the electroweak scale $m_{ew}$.
As discussed above,
the values of $\la t$ could be taken
of order one,
or, perhaps more plausibly,
of order of the Yukawa coupling $G_q$ for the
corresponding quark.

To gain intuition,
we have considered four scenarios,
each leading to a characteristic value of $|\de_P|$
as follows:
\bea
{\rm Scenario~0}:~&&~~~ \la t \sim 1~~,~~ m_l \sim m_{ew}~~,~~~~
|\de_P| \approx \fr {k m_{ew}^2}{a M_{Pl}}~~,~~~~
\nonumber\\
{\rm Scenario~1}:~&&~~~ \la t \sim G_{q_j}~~,~~ m_l \sim m_{ew}~~,~~~~
|\de_P| \approx \fr {G_q m_{ew}^2}{a M_{Pl}}~~,~~~~
\nonumber\\
{\rm Scenario~2}:~&&~~~ \la t \sim 1~~,~~ m_l \sim m_P~~,~~~~
|\de_P| \approx \fr {k m_P^2}{a M_{Pl}}~~,~~~~
\nonumber\\
{\rm Scenario~3}:~&&~~~ \la t \sim G_{q_j}~~,~~ m_l \sim m_P~~,~~~~
|\de_P| \approx \fr {G_q m_P^2}{a M_{Pl}}
\quad .
\label{ttt}
\eea
The parameter $k$ appearing in two of the scenarios
arises from the combination of the two $h_{q_j}$ factors
in Eq.\ \rf{t}.
Assuming no fine tuning,
$k$ is presumably of order one.
The factor $G_q$ refers to the heavier
of the two quarks in the $P$ meson,
which dominates the term $(h_{q_1} - h_{q_2})$ in Eq.\ \rf{t}.
Scenarios 0 and 3 are of lesser interest because
scenario 0 is probably experimentally excluded already
while scenario 3 results in CPT violation too small
to be detected in the foreseeable future.
The feature of interest is that intermediate scenarios
generate values of $\de_P$ comparable to the range
to be probed
in the next generation of experiments.
The intuition gained from these four candidate cases
suggests that
spontaneous CPT violation in a realistic string model
might generate experimentally measurable effects.

So far,
we have discussed only the effects
of string-inspired spontaneous CPT violation
in the $P$-$\overline P$ effective hamiltonian.
\it A priori, \rm
one might expect
additional CPT violating effects to appear
in ratios of decay amplitudes
along with those appearing directly in the $P$ mass matrix.
However,
in the string-inspired scenario
this is not the case because
there is no direct effect of the energy shifts
on the matrix elements of the decays.
Although contributions do appear from the interaction terms
of Eq.\ \rf{c}
when $\ps$ is $q_1$ and $\ch$ is $q_2$ or vice versa,
these effects are unobservable.
They can cause
decays that are forbidden in the standard model,
but with a suppression of at least two powers of
$m_l/M_{Pl}$.
Alternatively,
they can produce unobservable additional effects
in electroweak-type flavor-changing processes.
In principle,
an indirect effect could also appear
via energy-dependent
normalizations of the eigenstates,
but it too is unobservable because it is suppressed by at
least one power of the ratio
$m_l/M_{Pl}$.

\vglue 0.6cm
{\bf\noindent IV. STRINGY CPT VIOLATION IN MESON FACTORIES}
\vglue 0.4cm

We have seen in the previous section that stringy
CPT violation can generate nonzero
values of the parameters $\de_P$.
In the present section,
we consider the detection of such effects.
There are two issues involved for each $P$:
the expected signature of $\de_P$,
and the procedure for observing it.
Relatively good bounds on $\de_P$ can be obtained
in appropriate meson factories,
designed to generate high fluxes
of correlated $P_S$-$P_L$ pairs.
In sect. IVA,
we discuss the general framework of $P$ production
in meson factories
and present some formulae of use
in later sections.
Sects.\ IVB, IVC, and IVD consider in turn the cases
of $P\equiv K$, $B$, and $D$ mesons,
corresponding to $\ph$, $B$, and $\ta$-charm factories,
respectively.
For each case,
specifics of the signature of stringy CPT violation are discussed
and an analytical estimate is provided
relating the number of mesons produced
and the precision with which $\de_P$ is measured.
For these estimates,
we have exploited the elegant methods used
by Dunietz, Hauser, and Rosner
\cite{rosner}
for their analysis of the $K$ system in
the context of a $\ph$ factory.

\vglue 0.6cm
{\bf\noindent A. General Framework}
\vglue 0.4cm

The general idea behind a meson factory designed to
study properties of a $P$ meson is to generate
large numbers of a quarkonium state lying just above
the threshold for $P^0$-$\overline {P^0}$ production.
A state of this type strongly decays with a
relatively large branching ratio into correlated
$P^0$-$\overline {P^0}$
pairs.
Since $e^+$-$e^-$ machines generate virtual photons,
it is relatively easy to produce quarkonia
with $J^{PC}=1^{--}$.
For example,
for the case $P\equiv K$
the appropriate quarkonium is the $\ph$ meson,
while for $P\equiv B_d$ it is $\Up (4S)$,
for $P\equiv B_s$ it is $\Up (5S)$,
and for $P\equiv D$ it is $\ps (3770)$.

Since $C$ and $P$ are
conserved in strong interactions,
immediately after the strong decay of the quarkonium
the initial $P\overline P$ state $\ket i$ has
$J^{PC}= 1^{--}$.
For simplicity in what follows,
we work in the rest frame of the decay
and we choose the $z$ axis as the direction of the momenta
of the two $P$ mesons.
Denoting the mass eigenstates by
$\ket{P_S (\pm\z)}$ and $\ket{P_L (\pm\z)}$,
where $(+\z)$ means the particle is moving in the positive
$z$ direction and $(-\z)$ means it is moving in the
negative $z$ direction,
the normalized initial state $\ket i$ is given
by
\cite{lipkin}
\beq
\ket i = N \left[
\ket{P_S(\z)P_L(-\z)}
-\ket{P_L(\z)P_S(-\z)}
\right]
\quad ,
\label{iva}
\eeq
where the normalization constant
is $N^{-1} = \sqrt2 |1-\ep_P^2+\de_P^2|$.
The subsequent time evolution of the two $P$ mesons
is governed by the effective hamiltonian $\La$ of
sect.\ III:
\beq
\ket{P_S(t)}=e^{-im_St-\ga_St/2}\ket{P_S}~~,~~~~
\ket{P_L(t)}=e^{-im_Lt-\ga_Lt/2}\ket{P_L}
\quad ,
\label{ivb}
\eeq
which follows from Eq.\ \rf{n}.

Eventually,
the two $P$ mesons decay.
Let the meson moving in the positive $z$ direction
decay into the final state $\ket{f_1}$ at time $t_1$
while the other decays into $\ket{f_2}$ at $t_2$.
We measure the times $t_\al$, $\al=1,2$
in the rest frame of the quarkonium decay,
so they are given by the proper time of the
$P$ meson divided by the appropriate Lorentz gamma factor.
In what follows,
it is convenient
to define for each $\al$ the complex parameter
\beq
\et_\al
\equiv|\et_\al|e^{i\ph_\al}
= a_{\al L}/a_{\al S}
\quad ,
\label{ivc}
\eeq
as the ratio of the amplitude
$a_{\al L}=\bra{f_\al}T\ket{P_L}$
for the transition between $f_\al$ and $P_L$
to the amplitude
$a_{\al S}=\bra{f_\al}T\ket{P_S}$
for the transition between $f_\al$ and $P_S$.
Then,
the amplitude
${\cal{A}}_{12}(t_1, t_2)$ for the decay is
\beq
{\cal{A}}_{12}(t_1, t_2)
=N a_{1S}a_{2S}
\bigl (
 \eta_2e^{-i(m_St_1+m_Lt_2)-\half (\ga_St_1+\ga_Lt_2)}
-\eta_1e^{-i(m_Lt_1+m_St_2)-\half (\ga_Lt_1+\ga_St_2)}
\bigr ) .
\label{ivd}
\eeq
The interference between the two amplitudes
on the right-hand side is what makes possible the extraction
of information about CPT (and T) violation.
Note that
${\cal{A}}_{11}(t, t)=0$.

In sect.\ III, Eqs.\ \rf{oo} and \rf{ooo},
we defined $\De m$, $m$, $\De \ga$, and $\ga$.
It is useful to introduce in addition the quantities
\beq
\De \ph = \ph_1 - \ph_2
\label{ive}
\eeq
and
\beq
t=t_1+t_2~~,~~~~
\De t = t_2 - t_1
\quad .
\label{ivf}
\eeq
The decay rate $R_{12}(t,\De t)$
as a function of $t$ and $\De t$
can then be found:
\bea
R_{12}(t,\De t)
& = & |N a_{1S}a_{2S}|^2 e^{- \half \ga t}\nonumber\\
& &\quad\times \left [
|\eta_1|^2 e^{-\half \De \ga \De t}
+|\eta_2|^2 e^{\half \De \ga \De t}
- 2|\et_1\et_2|\cos(\De m\De t + \De \ph)
\right ] .
\label{ivg}
\eea

Experiments at meson factories will measure
integrated decay rates.
Following ref.\ \cite{rosner},
we consider first the once-integrated rates
for complete $t$ acceptance
\beq
I_{12}(\pm |\De t|) = \half
\int_{|\De t|}^\infty dt~
R_{12}(t,\pm |\De t|)
\quad ,
\label{ivh}
\eeq
where the time difference $\De t$ is kept constant.
The definition keeps separate track of the cases
where the decay into $f_1$ occurs before and after
that into $f_2$.
Writing $v= |\De t|$,
a short calculation gives
\beq
I_{12}(\pm v) =
\fr{|N a_{1S}a_{2S}\et_1|^2 }{\ga}e^{-\half \ga v}
\left [
e^{\mp\half \De \ga v}
+|r_{12}|^2 e^{\pm\half \De \ga v}
- 2|r_{12}|\cos(\De m v \pm \De \ph)
\right ]
\quad ,
\label{ivi}
\eeq
where we have defined
$r_{21} \equiv \et_2/\et_1 = |r_{21}| \exp (-i\De\ph)$.
Nonzero values of $\de_P$
(and $\ep_P$)
induce asymmetries between the two rates $I_{12}(\pm v)$,
entering through contributions to the $\et_\al$.

Typically,
a detector has limited $\De t$ acceptance,
lying in the range $\tau_1\le|\De t|\le\tau_2$,
say.
Define the twice-integrated rates
$\Ga_{12}^\pm(\ta_1,\ta_2)$
by
\beq
\Ga_{12}^\pm (\ta_1,\ta_2)=
\int_{\tau_1}^{\tau_2}dv~I_{12}(\pm v)
\quad .
\label{ivj}
\eeq
For most geometries
$\ta_1 = 0$.
We assume this in what follows,
and write $\ta_2 = \ta$.

The asymmetries induced by nonzero values of $\de_P$
(and $\ep_P$)
can be probed in terms of the rate asymmetry
$A_{12}(\ta)$,
defined as
\beq
A_{12}(\ta) =
\fr {\Ga_{12}^+(0,\ta)-\Ga_{12}^-(0,\ta)}
{\Ga_{12}^+(0,\ta)+\Ga_{12}^-(0,\ta)}
\quad .
\label{ivk}
\eeq
A calculation gives
\beq
A_{12}(\ta)=
\fr {(1-|r_{21}|^2)[h_S(\ta)-h_L(\ta)]
 -\fr 4 {b^2}\Im r_{21}~[\De m-be^{-\ga\ta/2}\sin(\De m\ta+\tilde\ph)]
}
    {(1+|r_{21}|^2)[h_S(\ta)+h_L(\ta)]
 -\fr 4 {b^2}\Re r_{21}~[\ga/2-be^{-\ga\ta/2}\cos(\De m\ta+\tilde\ph)]
} ,
\label{ivl}
\eeq
where
$h_{L(S)}(\ta)=(1-\exp(-\ga_{L(S)}\ta))/\ga_{L(S)}$,
and $b$ and $\tilde\ph$ are defined in Eq.\ \rf{ooo}.
In the subsequent sections,
we make particular use of the idealized asymmetry $A_{12}(\infty)$,
defined as the limit of $A_{12}(\ta)$
as $\ta\rightarrow\infty$:
\beq
A_{12}(\infty)=
\fr {
(1-|r_{21}|^2)(\ga_S^{-1}-\ga_L^{-1})
 -4\Im r_{21}~\De m/b^2
}{
(1+|r_{21}|^2)(\ga_S^{-1}+\ga_L^{-1})
-2\Re r_{21}~\ga/b^2
}
\quad .
\label{ivm}
\eeq
Introducing the conventional variables
\beq
x = \fr {2\De m}{\ga}~~,~~~~y = \fr{\De\ga}{\ga}
\quad ,
\label{ivn}
\eeq
we find
\beq
A_{12}(\infty)=
\fr {
2 x(1-y^2) \Im r_{21}
+y(1+x^2) (1-|r_{21}|^2)
}{
2(1-y^2) \Re r_{21}
-(1+x^2) (1+|r_{21}|^2)
}
\quad .
\label{ivo}
\eeq

\vglue 0.6cm
{\bf\noindent B. The $K\overline K$ System}
\vglue 0.4cm

In this subsection,
we consider the particular case $P\equiv K$,
for which the existence of CP violation has been
known for three decades
\cite{cronin}.
For this system,
experimental values of
$\ta_S = (8.922 \pm 0.020) \times 10^{-11}$ s,
$\ta_L = (5.17 \pm 0.04) \times 10^{-8}$ s,
and
$\De m = (3.522 \pm 0.016) \times 10^{-15}$ GeV
are available
\cite{PDG},
from which we calculate
\beq
\De \ga = (7.367 \pm 0.016) \times 10^{-15}~{\rm GeV} ~~,~~~~
\hat \ph = (43.71 \pm 0.14)^\circ
\quad .
\label{ivba}
\eeq
It follows that
\beq
x \approx y \simeq 1
\quad .
\label{ivbaa}
\eeq

A high intensity of correlated $K_S$-$K_L$ pairs
can be produced in a $\ph$ factory
\cite{lipkin,rosner}.
Several are now in the planning stages
or under construction,
including symmetric factories
(in which the $\ph$ are produced at rest)
at Frascati
\cite{daphne},
Novosibirsk
\cite{thompson},
and
KEK
\cite{kek},
and asymmetric ones
(in which the $\ph$ are produced with a boost)
at UCLA
\cite{cline}
and CEBAF (photoproduction)
\cite{dzierba}.
For electron-positron colliders,
typical peak luminosities
being discussed are of order
$10^{32}$ to $10^{33}$ cm$^{-2}$ s$^{-1}$.
The relevant cross section
is of order 4 $\mu$barn
corresponding to a peak rate of order
$10^2$ to $10^3~ \ph$ per second,
and one running year's worth of data
could produce up to about
$10^{11}$ $\ph$.
For photoproduction off protons,
the cross section is of order
0.3 $\mu$barn
and the peak rate of tagged photons
might be as high as $10^8$ or even $10^9$ s$^{-1}$,
with corresponding $\ph$ production rates
again of order $10^2$ to $10^3$ per second.

For this system,
the discussion of possible stringy spontaneous CPT violation
in section III specializes as follows.
Experimentally, $\De \ga\approx 2\De m $,
and $\hat\ph \simeq 45^\circ$.
Equation \rf{t}
therefore becomes
\beq
\de \approx \fr {\pm (h_s - h_d)} {\sqrt 2 \De m} e^{+3\pi i/4}
\quad .
\label{ivbb}
\eeq
Together with Eq.\ \rf{tt},
this shows that in the kaon system
stringy spontaneous CPT violation gives
$|\Im\de_K | \approx |\Re\de_K |$.
Plausible values of $|\de_K |$
encompass a range including ones within reach of the next
generation of experiments.

In the remainder of this subsection,
we combine the methods of
refs.\ \cite{barmin,TD,rosner}
with the formulae of section IVA
and some additional results
to estimate analytically
the level at which stringy spontaneous CPT violation
could be probed in a $\ph$ factory.
Since the branching ratios of the kaons
are dominated by the semileptonic and double-pion decays,
we focus on these cases.
Our analytical method provides a relatively quick means
of estimating the relevant limits.
A Monte-Carlo study of T- and CPT-violation tests
at a $\ph$ factory is presented in
ref.\ \cite{buch}.
A discussion of the possibility of measuring T, CPT,
and $\De S = \De Q$ violations in the kaon system
has recently been given
in ref.\ \cite{hayakawa}.
A useful collection of papers on CP violation
is given in
ref.\ \cite{jarlskog}.

Consider first the situation where the two final states
from the double-kaon decay are semileptonic:
$f_1=l^+\nu\pi^-$, $f_2=l^-\nu\pi^+$.
Assuming $\De S = \De Q$
and substituting Eq.\ \rf{j}
into the definition \rf{ivc}
gives
\beq
\eta_{l^\pm}=\pm 1-2\de_K+{\cal O}(\ep_K^2,\de_K^2)
\quad .
\label{ivbc}
\eeq
Using $\ga_L\ll\ga_S,\ga,\De m$
and taking
$\Re\de_K\not\ll\Im\de_K$
on the grounds of the string-inspired scenario,
it follows from Eq.\ \rf{ivl}
that
\beq
A_{l^+l^-}(\ta)=
{-4\Re\de_K~[h_S(\ta)-h_L(\ta)]+\fr 8 {b^2}\Im\de_K~
   [\De m-be^{-\ga\ta/2}\sin(\De m\ta+\tilde\ph)]
  \over
  h_L(\ta)+h_S(\ta)
   +\fr 2{b^2}[\ga/2-b\,e^{-\ga\ta/2}\cos(\De m\ta+\tilde\ph)]
 }
\label{ivbd}
\eeq
and so, using Eq.\ \rf{ivm} or \rf{ivo},
\beq
A_{l^+l^-}(\infty)\approx 4\Re\de_K
\quad .
\label{ivbe}
\eeq
Note that in the semileptonic case
the value of $\Im\de_K$
cannot be determined from the
limiting asymmetry with $\ta\rightarrow\infty$.
Instead,
one must fit directly to
Eq.\ \rf{ivbd} for finite $\ta={\cal O}(\ga_S^{-1})$.

Using the above information
together with experimental data on the branching ratios
of the semileptonic decays,
we can obtain the number of $\ph$ events needed to
detect a nonzero asymmetry at the $\pm N\si$ level,
where $\si$ is a standard deviation.
For an asymmetry $A=(N_+-N_-)/(N_++N_-)$,
the binomial distribution
implies that the expected number of events $\vev {N_+}$ required
to observe a nonzero $\vev{A}$ at the $N\si$ level is
$N^2(1+\vev{A})(1-\vev{A}^2)/(2\vev{A}^2)$.
In the present case,
this factor must be multiplied by
the inverse branching ratio for $\ph$ decaying
via two kaons into two final semileptonic states,
which is approximately 7800.
Using Eq.\ \rf{ivbe},
we find that the number $N_\ph$
of $\phi$ events needed to
reduce the error in $\Re\de_K$ to $\pm 1\si$ is
\beq
N_\ph(\Re\de_K) \simeq \fr {500}{\si^2}
\quad .
\label{ivbf}
\eeq
In this calculation,
the statistical accuracy
is improved by about a factor of two
by allowing the two charged leptons
in the final states to be any combination
of electrons and muons.
This is possible because the $\eta_{l^\pm}$
in Eq.\ \rf{ivbc} are independent of $l^\pm$
and because the decay-rate dependence
on $l^\pm$ cancels in the asymmetry.

We next consider double-pion final states,
which can provide information on $\Im\de_K$.
The parameters $\et_\al$
of Eq.\ \rf{ivc} for the case of $2\pi^0$ and $\pi^+\pi^-$
kaon decays are conventionally denoted as
$\et_{00}$ and $\et_{+-}$, respectively.
Present experimental values are
\cite{PDG}:
\bea
|\et_{+-} | = (2.268 \pm 0.024) \times 10^{-3}~~,&&~~~~
\ph_{+-} = (46.6\pm 1.2)^\circ
\nonumber\\
|\et_{00} | = (2.253 \pm 0.024) \times 10^{-3}~~,&&~~~~
\ph_{00} = (46.6\pm 2.0)^\circ
\quad .
\label{ivbg}
\eea
The standard theoretical calculation of these parameters
treats the
$|\De I | = \half$ and $|\De I | = \frac 3 2$
transitions as distinct,
and allows for isospin dependence
in any T and CPT violation
in the decay amplitudes.
A relatively tight bound on $|\Im\de_K |$ can be derived
\cite{barmin,TD}
from these values
provided one neglects contributions to the effective hamiltonian
coming from $3\pi$ decays and
from violations of the $\De S = \De Q$
rule in semileptonic decays.
Introducing
$\ph_\ep \approx (2\ph_{+-} + \ph_{00})/3$,
we find
\bea
|\Im\de_K |&\approx& |\et_{+-}|~|\cos\hat\ph \sin (\ph_\ep-\hat\ph) |
\nonumber\\
&=&(8.3\pm 2.9)\times10^{-5}
\quad .
\label{ivbh}
\eea
A completely reliable bound taking into account the effects
mentioned is less constrained
\cite{lavoura1} because then $39.5^\circ < \ph_\ep < 47.4^\circ$,
from which we find
$|\Im\de_K | < 1.3\times10^{-4}$.
In either case,
the real part of $\de_K $ is not as tightly bounded
because in $2\pi$ decays
it always appears in conjunction with CPT-violating
pieces of the decay amplitudes.
The discussion in section III implies these are zero
for stringy spontaneous CPT violation,
in which case we can apply the condition
$|\Re\de_K | \approx |\Im\de_K |$.

The error in $|\Im\de_K |$ in Eq.\ \rf{ivbh} is dominated by
that in $\ph_\ep$.
It can therefore be decreased by improving the precision
with which $\ph_{+-}$ and $\ph_{00}$ are measured.
Their difference $\De\ph$
can be determined from Eq.\ \rf{ivi},
for example,
by taking as the two final states
$\pi^+\pi^-$ and $\pi^0\pi^0$
and fitting for finite values of $\De t$.
The precision in the value for $\De\ph$ thus obtained
can be estimated analytically using
Eq.\ \rf{errorA} from the appendix,
where the two functions $f(t)$, $g(t)$
and the amplitude $\al$ are taken as
\bea
f(v)&=&e^{-\ga_Sv}+e^{-\ga_Lv}
-2 e^{-\ga v/2}\cos\De mv~\cos\De\ph
\nonumber\\
g(v)&=&e^{-\ga v/2}\sin\De mv
\nonumber\\
\al &=& 2\sin\De\ph
\quad .
\label{ivbj}
\eea
Combining the result of this calculation
with the appropriate inverse branching ratio for the
double-$2\pi$ decay of the $\ph$,
which is approximately $4800$,
we obtain the number $N_\ph$ of $\ph$ events needed to
reduce the error in $\De\ph$ in degrees
to within $\pm 1\si$ as
\beq
N_\ph(\De\ph) \simeq
\fr {3.2\times 10^{9}}{\si^2}
\quad .
\label{ivbk}
\eeq

Determining $\ph_{+-}$ or $\ph_{00}$ separately is less
straightforward.
A relatively good approach is to look at final states
consisting of $f_1 \equiv 2\pi$ with appropriately
charged pions
and $f_2\equiv \pi^\mp l^\pm \nu$,
for which $\et_{l^\pm}=\pm 1-2\de_K\simeq \pm 1$.
For example, using Eq.\ \rf{ivi} with $\De t>0$
permits $\ph_{+-}$ or $\ph_{00}$
to be determined from the interference term,
which carries opposite sign for opposite lepton charges.
We use the method
described in the second part of the appendix,
which provides a means to estimate analytically
the errors in asymmetries
consisting of linear combinations of known functions.
Taking $\al=2|\et_{\pi\pi}|$
and $\la\approx \be =\De m$,
we find from Eq.\ \rf{errorch}
that a measurement in degrees of $\ph_{\ep}$
to an accuracy of $\pm 1\si$ requires a number $N_\ph$ of $\ph$
given by
\beq
N_\ph (\ph_{\ep}) \simeq
\fr {1.3\times 10^{12}}{\si^2}
\quad .
\label{ivbl}
\eeq
In this estimate,
we summed over both electron and muon contributions,
obtaining a combined inverse branching ratio
of approximately $2700$.
Treating the error in $|\et_{+-} |$ and $\hat\ph$
as negligible,
the result \rf{ivbl} implies that the number
$N_\ph$ of $\ph$ events needed to reduce the error
in $\Im \de_K$ to $\pm 1 \si$ is
\beq
N_\ph (\Im\de_K) \simeq
\fr {1100}{\si^2}
\quad .
\label{ivbm}
\eeq

\vglue 0.6cm
{\bf\noindent C. The $B\overline {B}$ System}
\vglue 0.4cm

We next examine the $P$-$\overline P$ system
with $P$ taken as one of the two neutral $B$ mesons,
$B_d$ or $B_s$.
We primarily focus on the case where $P\equiv B_d$.
Towards the end of this section we comment
on the other choice.

The experimental data for
$B_d\overline{B}_d$ system are
\cite{PDG}
$\bar\ta_B = (12.9\pm 0.5)\times 10^{-13}$ s
and $|\De m | = (3.6 \pm 0.7) \times 10^{-13}$ GeV,
which implies
$|x| = 0.71 \pm 0.14$.
However,
these values have been derived
under the assumption that CPT is preserved.
If this assumption is relaxed,
the experimental value of $x$ becomes a lower bound
\cite{kobayashi},
\beq
|x| \ge 0.71 \pm 0.14
\quad .
\label{ivca}
\eeq
We use this bound in what follows.

At present,
an experimental value of $y$ is unavailable.
On theoretical grounds,
perturbative calculations via the box diagram
are expected to provide
an accurate estimate of $y$ for this system
because the dominant intermediate states
are the top and charm quarks and so
short-distance effects should dominate over
dispersive ones.
For details of these calculations and a guide to the
large literature,
the reader is referred to any of the standard
reviews on this subject;
see, for example,
refs.\ \cite{chau,franzini,paschos,bigi}.
The results of these calculations suggest
\beq
|y| \simeq O(10^{-2}) \ll |x|
\quad .
\label{ivcb}
\eeq

A relatively large number
of correlated $B_d$-$\overline{B}_d$
pairs can be generated in a $B$ factory,
which provides an intense source of $\Up (4S)$.
Discussions of some of the many proposals now being
considered can be found in
ref.\ \cite{beauty93}.
In particular,
Cornell and SLAC in the U.S. and KEK in Japan
are proceeding with asymmetric $B$ factories.
Peak luminosities anticipated are in the range
$10^{33}$ to $10^{34}$ cm$^{-2}$ s$^{-1}$.
The cross section for $\Up (4S)$ production in
electron-positron machines is of order 1.2 nanobarn,
so peak production could lie in the range
one to 10 Hz.
A running year therefore could provide about $10^7$ to $10^8$
correlated $B_d$-$\overline{B}_d$ pairs.

The discussion in section III
of potential stringy spontaneous CPT violation
specializes for this system
as follows.
Equation \rf{ivcb} implies $\De m \gg \De \ga$,
so $\hat\ph \simeq \pm \pi/2$ and Eq.\ \rf{t}
becomes
\beq
\de_{B_d} \approx \pm \fr {h_b}{\De m}
\quad .
\label{ivcc}
\eeq
It follows from this or Eq.\ \rf{tt} that
stringy spontaneous CPT violation predicts
$|\Re \de_{B_d}| \gg |\Im \de_{B_d}|$
in the $B_d$ system,
with magnitude potentially accessible to experiment.

We next discuss briefly the issue of measuring
$\de_{B_d}$
experimentally in a $B$ factory.
Consider the case in which the $\Up (4S)$
produces a double-semileptonic final state,
$f_1=l^+\nu\pi^-$, $f_2=l^-\nu\pi^+$.
At present,
it is not known whether
T and CPT violation are large, small, or zero
in this system.
If the violation is large,
the integrated asymmetry
\rf{ivm} or \rf{ivo} provides a relation between
the real and imaginary parts of
the ratio $r_{21}$.

If, however, the violation is small,
the ratio $r_{21}$ can be related to $\de_{B_d}$
using an expression analogous to Eq.\ \rf{ivbc}.
Then,
Eqs.\ \rf{ivm} and \rf{ivo}
reduce to
\bea
A_{l^+l^-}(\infty)
&\approx&
{
-4\Re\de_{B_d}(\ga_S^{-1}-\ga_L^{-1})+8\Im\de_{B_d}\De m/b^2
\over
 (\ga_L^{-1}+\ga_S^{-1})+\ga/b^2
}
\nonumber\\
&\approx&
4~\fr{
x\Im \de_{B_d}
+y(1+x^2) \Re \de_{B_d}
}{
2 + x^2
}
\quad .
\label{ivcd}
\eea
A measurement of this asymmetry
therefore provides one probe of
$\de_{B_d}$.

For stringy spontaneous CPT violation
Eq.\ \rf{tt} gives
\beq
y~\Re \de_{B_d} \approx \pm x~\Im \de_{B_d}
\quad ,
\label{ivcda}
\eeq
and Eq.\ \rf{ivcd} reduces to
\beq
A_{l^+l^-}(\infty)
\approx
\pm\fr {4x(1+x^2\pm 1)}{2+x^2}~\Im\de_{B_d}
\quad .
\label{ivce}
\eeq
Taking $|x| \simeq 0.71$ gives
\beq
|A_{l^+l^-}(\infty)|
\approx k~|\Im\de_{B_d}|
\quad ,
\label{ivcea}
\eeq
where $k\simeq \half$ or $k\simeq 3$,
depending on the choice of sign in Eq.\ \rf{ivce}.

To improve statistical accuracy,
we can again take advantage of the fact that
several decay channels have equal
values of $\et_{l^\pm}=\pm 1-2\de_{B_d}$.
These include semileptonic decays,
summing to a branching ratio of about 10\%,
along with any other channels
in $B_d$ decay
forbidden to a good approximation
in $\overline{B}_d$ decay,
paired with the corresponding CP conjugates.
In principle,
this includes a total branching ratio of at least
about 20\% and perhaps much larger.
Taking this into account and using Eq.\ \rf{ivce}
with $k\simeq O(1)$,
we find that the number
$N_{\Up(4S)}$
of events needed to reduce the error in $\Im\de_{B_d}$ to
within $\pm 1 \si$ is
no more than about
\beq
N_{\Up(4S)}(\Im\de_{B_d})
\simeq \fr {5}{\si^2}
\quad .
\label{ivcf}
\eeq

It would also be desirable
to distinguish experimentally
the case of stringy spontaneous CPT
violation from a generic scenario.
If, for example,
one allows for values of
$\Im\de_{B_d}$
comparable to those of $\Re\de_{B_d}$,
the asymmetry \rf{ivcd} reduces instead to
\beq
A_{l^+l^-}(\infty) \approx
\fr {4x} {2+x^2}~
\Im\de_{B_d}
\simeq
\Im\de_{B_d}
\quad .
\label{ivcg}
\eeq
Although we are unaware of any explicit
theoretical mechanism that could produce
this result,
it nonetheless is not experimentally excluded.
It would therefore appear desirable to have an alternative
independent means of measuring $\de_{B_d}$.
The resolution of this issue
depends on successfully disentangling
different CP-violation effects
and therefore awaits a complete study of T and CPT violation
in the $B$ system.\footnote{
Since the completion of the present work,
some progress along these lines has been made;
see ref.\ \cite{ck}.}

So far,
our focus has been on the case $P\equiv B_d$.
In the remainder of this subsection,
we consider the alternative choice $P\equiv B_s$.

The experimental data for the key parameters
in the $B_s$-$\overline{B}_s$
system are limited at present,
so we content ourselves with theoretical predictions.
See, for example,
\cite{franzini,paschos,bigi}
and references therein.
On the basis of current knowledge of the CKM matrix
and the $B_d$ system,
it is expected that
\bea
|x_s| &\gsim & 10
\quad ,
\nonumber\\
|y_s| &\simeq & 2\times10^{-2}
\quad .
\label{ivcca}
\eea
The inequality again reflects the uncertainty
in $x$ arising from the possibility of CPT
violation,
as in Eq.\ \rf{ivca} above.
The theoretical uncertainties in these numbers
are probably at least a factor of two.

The system might be explored in a $B$ factory
focusing on the production of $\Up (5S)$.
However, both the cross section for production
and the branching ratio for
$B_s$-$\overline{B}_s$
decay are smaller than in the $B_d$ case,
so a luminosity of perhaps two orders of magnitude
greater than that in the planned $B$ factories
would be required to produce
comparable numbers of correlated
$B_s$-$\overline{B}_s$ pairs.

Anticipated effects arising from
possible stringy spontaneous CPT violation
are analogous to the $B_d$ system.
Thus,
we obtain
\beq
\de_{B_s} \approx \de_{B_d} \approx \pm \fr {h_b}{\De m}
\quad ,
\label{ivccb}
\eeq
and $\hat\ph \simeq \pi/2$ again, so
$|\Re \de_{B_s}| \gg |\Im \de_{B_s}| $
in the $B_s$ system too.

Some differences between the two systems
arise because $|x_s|> |x|$.
The implications of this for experiments
can be found using the methods discussed above.
For example,
the string scenario implies that
the double-semileptonic asymmetry
Eq.\ \rf{ivcd} reduces to
\beq
A_{l^+l^-}(\infty)
\approx
\pm 4 x_s~\Im\de_{B_s}
\gsim \pm 40~\Im\de_{B_s}
\quad .
\label{ivccc}
\eeq
This theoretical result suggests that
a measurement of $\Im\de_{B_s}$ is about three orders
of magnitude more favorable
than the $B_d$ case;
cf.\ Eq.\ \rf{ivcf}.
However, as mentioned above,
the experiment may be
two or more orders of magnitude harder
because of the sizes of the associated
production cross sections and branching ratios.

\vglue 0.6cm
{\bf\noindent D. The $D \overline D$ System}
\vglue 0.4cm

In this subsection,
we consider the case $P\equiv D$.
At present, there is no experimental evidence for mixing
in the $D$-$\overline{D}$ system.
This issue could be examined to a relatively high
degree of precision
in a $\ta$-charm factory
such as that originally envisaged in Spain
\cite{kirkby}.
With a design luminosity peaking
between $10^{32}$ and $10^{33}$ cm$^{-2}$ s$^{-1}$,
a machine operating at the $\ps (3700)$
could generate yearly between $10^7$ and $10^8$
correlated $D$-$\overline{D}$ pairs.

Calculations of short-distance effects
via the box diagram suggest only very small effects,
giving $|x| \approx |y| \simeq 10^{-6}$.
Summaries of the situation and
a guide to the relevant literature can be found in
\cite{chau,paschos}.
In addition to the short-distance effects,
long-range or dispersive contributions
to $x$ and $y$ must be considered.
These are dominated by intermediate lower-mass
states and are difficult to estimate with precision.
In the kaon system,
the short- and long-range contributions
are believed to be comparable
\cite{wolfenstein1,hill,donoghue1}.
In the $B$ systems the short-range
contributions dominate, as discussed above.
In contrast,
the $D$ system is believed to be dominated
by dispersive effects
\cite{wolfenstein2,donoghue2}.
These generate values of $|x|$ and $|y|$
larger than the box contributions by several
orders of magnitude.
In what follows,
we take for illustrative purposes $\De m \approx \De \ga$
such that
\cite{wolfenstein2}
\beq
|x| \approx 2|y| \simeq 2 \times 10^{-2}
\quad .
\label{ivda}
\eeq

With this assumption,
Eq.\ \rf{t} of section III becomes
\beq
\de \approx \fr {\pm i \sqrt{2}h_c} {\sqrt 5 \De m} e^{i\hat\ph}
\quad ,
\label{ivdb}
\eeq
with $\hat\ph \simeq 63^\circ$.
The relationship corresponding to Eq.\ \rf{tt} is
$2|\Im\de_D | \approx |\Re\de_D |$.
Plausible values of $|\de_D |$
again encompass a range within experimental reach.

Information about $\de_D$ could be obtained
from a measurement of the double-semileptonic asymmetry
of Eq.\ \rf{ivm} or \rf{ivo}.
Assuming T and CPT violation are small,
as usual,
Eq.\ \rf{ivcd} holds except with $\de_B$ replaced by $\de_D$.
The largest asymmetry here appears when
the signs are such that a term linear in $x$ survives,
whereupon we find
\beq
A_{l^+l^-}(\infty)
\approx 4x~ \Im \de_D
\simeq \pm 10^{-1}~\Im \de_D
\quad .
\label{ivdc}
\eeq
In this case,
the asymmetry provides a measurement of
$\Im\de_D$ rather than the real part,
as occurred in the kaon system.
This arises even though
the mixing parameters obey $|x|\approx |y|$
in both cases because
for $K$ the mixing parameters are
of order one whereas for $D$ they are much less than one.

Branching ratios for the $D$ meson are better known than
for the $B$ system.
We can again use a combination of semileptonic and
other decay channels with
$\et_{l^\pm}=\pm 1-2\de_D$,
which here includes a total branching ratio of at least
50\%.
Assuming Eq.\ \rf{ivdc},
we obtain the number
$N_{\ps(3770)}$
of events needed to reduce the error in $\Im\de_{D}$ to
within $\pm 1 \si$ as approximately
\beq
N_{\ps(3770)}(\Im\de_D)
\simeq \fr {200}{\si^2}
\quad .
\label{ivdd}
\eeq
In principle,
further information about $\de_D$ is encoded in asymmetries
involving non-semileptonic final states.
We do not pursue this issue in the present work.

\vglue 0.6cm
{\bf\noindent V. SUMMARY}
\vglue 0.4cm

In this paper,
we presented a generic theoretical framework
for possible stringy spontaneous CPT violation
and examined some experimental consequences.
We concentrated on the detection of
effects using a sensitive tool:
the $P \overline P$ interferometers.
We found that a range of string-inspired values
of the CPT-violating parameter $\de_P$
in these systems
can be explored in suitable meson factories.
For the case $P\equiv K$,
the string scenario gives
$|\Re \de_K | \approx |\Im \de_K |$.
For $P\equiv B_d$ or $B_s$,
$|\Re \de_B | \gg |\Im \de_B | $,
while for $P\equiv D$,
a rough estimate suggests
$|\Re \de_K | \approx 2|\Im \de_K |$.
Estimates were given of the number of events
needed to reduce the error in the real or imaginary
part of $\de_P$ to within one standard deviation.
These are presented
in Eqs.\ \rf{ivbf} and \rf{ivbm} for the case $P\equiv K$,
in Eq.\ \rf{ivcf} for $P\equiv B_d$,
and in Eq.\ \rf{ivdd} for $P\equiv D$.
The case $P\equiv B_s$ was also briefly discussed.
The anticipated mesons containing the top quark,
$T_u$ and $T_c$,
were not considered
since it is likely the top quark decay proceeds
too rapidly to permit hadronization.
Effects on other physics,
including for example the lepton sector,
may also exist and are under investigation.
Note, however,
that although the stringy mechanism
might also lead to violation of other discrete symmetries,
any effects are likely to be masked by violations
appearing from known effects in particle field theories.

One point that emerges from this work is that
detecting stringy effects,
while certainly difficult,
is not necessarily impossible.
Although the absence of a satisfactory
and realistic string theory
precludes definitive predictions for low-energy effects,
several possibilities do exist.
Among them is the
stringy spontaneous CPT violation
discussed here.
We have seen that,
if it does occur,
the violation
appears in a particular sector
and can have magnitude
accessible in the next generation of experiments.

\vglue 0.6cm
{\bf\noindent VI. ACKNOWLEDGMENTS}
\vglue 0.4cm

We thank Stuart Samuel for useful comments.
V.A.K.\ has benefitted
from the hospitality of the Aspen Center for Physics
in the summers of 1992 and 1993,
when parts of this work were performed.
R.P.\ thanks the high-energy physics group
at Indiana University for hospitality during
an extended visit.
This work was supported in part
by the United States Department of Energy
under grant number DE-FG02-91ER40661.

\vfill\eject
{\bf\noindent APPENDIX}
\vglue 0.4cm

Suppose we want to measure an asymmetry in a decay curve,
obtained as a function of the elapsed time
$t$ since production in the inertial frame of the decay.
Let there be two possible decay products,
denoted by $+$ and $-$,
occurring with probabilities $f_+(t)$ and $f_-(t)$.
For the cases of interest in the text,
we have
\beq
f_\pm(t)=f(t)\pm \al g(t)
\label{solveA}
\eeq
where $f$ and $g$ are known functions
and $\al $ is a constant that is the focus of the
experimental measurement.
We take the asymmetry $A(t)$ to be defined by
\beq
A(t)\equiv \al g(t) =\half (f_+(t) - f_-(t))
\quad .
\label{asymmetry}
\eeq

Let the experiment involve values of $t$ lying in the range
$t\in [a,b]$.
First,
we divide the interval in bins.
After $N$ measurements,
the number of $\pm$ events in bin $n$ is
\beq
N_\pm(n)\approx Cf_\pm(t_n)~~,~~~~
C=\fr N{\sum_n [f_+(t_n)+f_-(t_n)]}
\eeq
where $C$ is a normalization constant.
We can use the form of Eq.\ \rf{solveA}
to find an expression for $\al$
in terms of a function
$h(t_n)$ to be determined below:
\beq
\al ={\sum_n A(t_n) h(t_n)\over
\sum_ng(t_n)h(t_n)}
\quad .
\label{getA}
\eeq

The error in the measured value for the asymmetry
$A(t_n) =[N_+(n)-N_-(n)]/2C$ is
$\sqrt{N_+(n)+N_-(n)}/2C\approx\sqrt{f(t_n)/2C}$.
Thus,
using Eq.\ \rf{getA} to obtain $\al $,
the resulting error in $\al$ is
\bea
\delta_h\al
&=&
(2C)^{-\half}\fr {
[\sum_n h(t_n)^2f(t_n)]^\half
}{
\sum_ng(t_n)h(t_n)
}
\nonumber\\
&=&
(2N)^{-\half}
\fr {
[\sum_n f(t_n)]^\half
[\sum_n h(t_n)^2f(t_n)]^\half
}{
\sum_n g(t_n)h(t_n)
}
\quad ,
\eea
or, taking the continuum limit,
\beq
\delta_h\al =
(2N)^{-\half}
\fr {
[\int f(t)dt]^\half[\int h(t)^2f(t)dt]^\half
}{
\int h(t)g(t)dt}
\quad .
\eeq
At this point we can determine $h(t)$ by requiring that the resulting
error in $\al $ be minimal.
Thus,
\beq
{\de\over\de h(t)}\left[
{\sqrt{\int h^2f dt}\over\int hg \,dt}
\right] =0
\quad .
\eeq
This yields
\beq
h(t)\propto g(t)f(t)^{-1}
\quad ,
\eeq
resulting in an error
\beq
\de \al =(2N)^{-\half}\sqrt{\int f\,dt\over \int g^2f^{-1}dt}
\quad .
\label{errorA}
\eeq

This result can be easily extended to a
situation where the asymmetry
is a linear combination of known functions:
\beq
A(t_n)=\sum_j\al _jg_j(t)
\quad .
\eeq
In this case,
we introduce functions $h_i(t)$,
chosen to satisfy
\beq
\sum_ng_j(t_n)h_i(t_n)\propto\de_{ij}
\quad ,
\label{ortho}
\eeq
so that the quantities $\al _i$ are determined by
\beq
\al _i={\sum_n A(t_n) h_i(t_n)\over
\sum_ng_i(t_n)h_i(t_n)}
\quad .
\eeq
Minimizing the error,
we find that the $h_i$ are proportional to
$g_if^{-1}$.
Then,
Eq.\ \rf{ortho}
implies that the $g_i$ must
satisfy the orthogonality conditions
\beq
\sum_n{g_i(t_n)g_j(t_n)\over f(t_n)}=0~~,~~~~i\ne j.
\label{orthog}
\eeq
If the chosen functions $g_i$ do not satisfy these conditions,
the Gramm-Schmidt orthogonalization procedure can be applied.
We finally obtain the induced error in the parameters $\al_i$:
\beq
\de \al _i=(2N)^{-\half}\sqrt{\int f\,dt\over \int g_i^2f^{-1}dt}
\quad .
\label{errorAi}
\eeq

In the text,
we are interested in applying this method to a case
with
\bea
f(t)&=&e^{-\la t}\\
A(t)&=&\al e^{-\la t}\cos(\be t-\ch)
\quad ,
\eea
where
$\la$ and $\be $ are known parameters,
while $\al $ and
$\ch$ are to be measured experimentally.
We are particularly interested in the
induced error in $\ch$.
Here,
the antisymmetry is \it not \rm
in the form of a linear combination
of basis functions.
However,
this can be remedied by setting
$\cos(\be t-\ch)=\cos \be t~\cos \ch +\sin \be t~\sin \ch $
and using Gramm-Schmidt orthogonalization
on the functions $e^{-\la t}\cos \be t$
and $e^{-\la t}\sin \be t$.
We thereby find that the functions
\bea
g_1(t)&=&e^{-\la t}\cos \be t\\
g_2(t)&=&e^{-\la t}\left (\sin \be t-{w\over2+w^2}\cos \be t\right)
\quad ,
\eea
where $w\equiv \la/\be$,
satisfy the condition \rf{orthog}.
We can then directly apply
\rf{errorAi} to obtain the errors
in the coefficients $\al _1$, $\al _2$.
For the error in $\ch=\arctan(\al _2/\al _1)$,
we obtain
\beq
\de \ch=(2N)^{-\half}\al ^{-1}(4+w^2)^\half
\left [{\cos^2\ch\over2+w^2}+
       \half\left(\sin\ch+{w\cos\ch\over2+w^2}\right)^2
       \right ]^\half
\quad .
\label{errorch}
\eeq

\vglue 0.6cm
{\bf\noindent REFERENCES}
\vglue 0.4cm

\end{document}